\documentstyle{mn}
\newif\ifAMStwofonts

\input epsf

\ifoldfss
  \ifCUPmtlplainloaded \else
    \NewTextAlphabet{textbfit} {cmbxti10} {}
    \NewTextAlphabet{textbfss} {cmssbx10} {}
    \NewMathAlphabet{mathbfit} {cmbxti10} {} 
    \NewMathAlphabet{mathbfss} {cmssbx10} {} 
  \fi
  \ifAMStwofonts
    \ifCUPmtlplainloaded \else
      \NewSymbolFont{upmath} {eurm10}
      \NewSymbolFont{AMSa} {msam10}
      \NewMathSymbol{\upi}     {0}{upmath}{19}
      \NewMathSymbol{\umu}     {0}{upmath}{16}
      \NewMathSymbol{\upartial}{0}{upmath}{40}
      \NewMathSymbol{\leqslant}{3}{AMSa}{36}
      \NewMathSymbol{\geqslant}{3}{AMSa}{3E}

    \fi
  \fi
\fi 

\ifnfssone
  \newmathalphabet{\mathit}
  \addtoversion{normal}{\mathit}{cmr}{m}{it}
  \addtoversion{bold}{\mathit}{cmr}{bx}{it}
  \newmathalphabet{\mathbfit} 
  \addtoversion{normal}{\mathbfit}{cmr}{bx}{it}
  \addtoversion{bold}{\mathbfit}{cmr}{bx}{it}
  \newmathalphabet{\mathbfss} 
  \addtoversion{normal}{\mathbfss}{cmss}{bx}{n}
  \addtoversion{bold}{\mathbfss}{cmss}{bx}{n}
  \ifAMStwofonts
    \ifCUPmtlplainloaded \else
      \UseAMStwoboldmath
      \makeatletter
      \new@mathgroup\upmath@group
      \define@mathgroup\mv@normal\upmath@group{eur}{m}{n}
      \define@mathgroup\mv@bold\upmath@group{eur}{b}{n}
      \edef\UPM{\hexnumber\upmath@group}
      \new@mathgroup\amsa@group
      \define@mathgroup\mv@normal\amsa@group{msa}{m}{n}
      \define@mathgroup\mv@bold\amsa@group{msa}{m}{n}
      \edef\AMSa{\hexnumber\amsa@group}
      \makeatother
      \mathchardef\upi="0\UPM19
      \mathchardef\umu="0\UPM16
      \mathchardef\upartial="0\UPM40
      \mathchardef\leqslant="3\AMSa36
      \mathchardef\geqslant="3\AMSa3E
    \fi
  \fi
\fi 

\ifnfsstwo
  \DeclareMathAlphabet{\mathbfit}{OT1}{cmr}{bx}{it}
  \SetMathAlphabet\mathbfit{bold}{OT1}{cmr}{bx}{it}
  \DeclareMathAlphabet{\mathbfss}{OT1}{cmss}{bx}{n}
  \SetMathAlphabet\mathbfss{bold}{OT1}{cmss}{bx}{n}
  \ifAMStwofonts
    \ifCUPmtlplainloaded \else
      \DeclareSymbolFont{UPM}{U}{eur}{m}{n}
      \SetSymbolFont{UPM}{bold}{U}{eur}{b}{n}
      \DeclareSymbolFont{AMSa}{U}{msa}{m}{n}
      \DeclareMathSymbol{\upi}{0}{UPM}{"19}
      \DeclareMathSymbol{\umu}{0}{UPM}{"16}
      \DeclareMathSymbol{\upartial}{0}{UPM}{"40}
      \DeclareMathSymbol{\leqslant}{3}{AMSa}{"36}
      \DeclareMathSymbol{\geqslant}{3}{AMSa}{"3E}
    \fi
  \fi
\fi 

\ifCUPmtlplainloaded \else
  \ifAMStwofonts \else 
    \def\upi{\pi}
    \def\umu{\mu}
    \def\upartial{\partial}
  \fi
\fi

\title{Globular Cluster System Erosion 
and Nucleus Formation in Elliptical Galaxies}
\author[R.~Capuzzo--Dolcetta and A.~Tesseri]
       {R.~Capuzzo--Dolcetta,$^1$  A.~Tesseri,$^1$\\
        $^1$ Istituto Astronomico, Universit\`a La Sapienza,\\
 Via G. M. Lancisi 29, I-00161, Roma, Italy\\
dolcetta@astrmb.rm.astro.it,
tesseri@astrmb.rm.astro.it}

\date{Accepted .
      Received ;
      in original form }

\pagerange{\pageref{firstpage}--\pageref{lastpage}}
\pubyear{1999}

\begin{document}

\maketitle

\label{firstpage}

\begin{abstract}
The radial distribution of globular clusters in galaxies is always less
peaked to the centre than the halo stars'. Extending previous work to a sample
of HST globular cluster systems in ellipticals, we evaluate  the 
number of clusters lost to the galactic centre as the integrals
of the difference between the observed globular cluster system distribution and
the underlying halo light profile.
It results that the initial populations of globular clusters were from
 $25\%$ to $50\%$ richer than now.
This significant number of \lq missing \rq~ globular clusters supports the 
hypothesis that a large quantity of globular cluster mass in form of globular clusters decayed and destroyed has been lost to the
galactic centres,  where plausibly contributed to formation and feeding of a massive object therein.
It is relevant noting that the observed correlation between the core radius of  the globular
cluster system and the  parent galaxy luminosity can be interpreted
as a result of evolution.
\end{abstract}


\section{Introduction}
It is a quite accepted statement that globular cluster systems
(GCSs)
in galaxies are less concentrated to the centre than the bulge component, even though in some cases the available data are not
 good enough to compare reliably the cluster and halo distributions.

Two clear examples of galaxies where bulge star distributions are more centrally peaked than GCSs are the two Virgo giant galaxies
M87 and M49 (Harris, 1986). Also our Galaxy and 
M31 show a similar behaviour (Capuzzo--Dolcetta \& Vignola, 
1997); no case has been found yet where the GCS is more centrally concentrated than the halo (Harris, 1991). 
There is a general agreement on that the difference in the profiles is real, and not
due to observational bias.
The difference in the spatial profiles is confirmed by the recent HST WFPC2 observations of 14 elliptical galaxies with kinematically distinct cores (Forbes et al. 1996).
  Deep discussions of  the possible explanations of this fact
can be found in Capuzzo--Dolcetta \& Vignola (1997) and Capuzzo--Dolcetta \& Tesseri (1997) (hereafter CDV and CDT), and are not worthly
repeated here.
\par\noindent We just say that two are the likely  possibilities: 

$i)$ the distributions reflect a difference in the formation ages of halo stars and globular clusters, as suggested by Harris \& Racine (1979) and by Racine (1991); 

$ii)$ the difference is due to evolution of the GCS distribution caused by dynamical friction and tidal interaction with a massive galactic nucleus (Ostriker, Binney \& Saha, 1989; Capuzzo--Dolcetta, 1993; CDV).

While it is still hard to state firmly which is the 
correct explanation (we suspect that velocity data will be needed), the role of
dynamical effects in modifying the initial density profile
of a GCS has been quantitatively ascertained.

Actually, CDT showed clearly how a GCS 
in a triaxial galaxy evolves in such a way that its radial
profile gets significantly flatter in the central regions 
than the initial. This is exactly what observed in galaxies.

Of course the evolution of a GCS may have had important
consequences for its mother galaxy; for example, a significant
amount of mass in form of dynamically decayed clusters
may have been carried to the very inner galactic regions where it
is swallowed by a nucleus which, consequently, increases in size.
This hypothesis, qualitatively raised first by Ostriker, Binney \& Saha (1989),
has been subsequently checked and confirmed by Capuzzo--Dolcetta (1993).

On an observational side, Mc Laughlin (1995) and  CDV, scaling the presently observed radial profiles of GCSs of some galaxies (the Milky Way, M31 and M87) to those of their haloes, were able to evaluate the integral of the difference between 
radial profiles 
giving numbers that are interpreted as the numbers of cluster disappeared.
The hypothesis that the shape of the distributions of the GCS and the bulge were initially the same was done.
Proceeding in this way, CDV found it plausible that compact nuclei in our Galaxy, M 31 and M 87 have accreted many decayed globulars in the first few Gyrs of life.

The purpose of the present work is the extension of the work
by Mc Laughlin and CDV to the set of 
HST globular cluster data by Forbes et al. (1996) which 
constitutes a good sample of homogeneous data.
\begin{table}
\caption{Correlation coefficients (r) and loss function (l) for the analytical
fits to the GCS (core model) and galaxy light (core model and de Vaucouleurs' model) distributions.}
\label{symbols}
\begin{tabular}{@{}lcccc}
Galaxy & r$_{GCS}$ & l$_{GCS}$ & l$_{gal}$(core) & l$_{gal}$(de V.) \\
\hline
  NGC 4365 & 0.984 & 0.07 & 0.03 & 0.04 \\
  NGC 4494 & 0.999 & 0.03 & 0.04 & 0.04 \\
  NGC 1700 & 0.787 & 0.27 & 0.01 & 0.004 \\
  NGC 5322 & 0.998 & 0.05 & 0.02 & 0.04 \\
  NGC 5982 & 0.972 & 0.10 & 0.01 & 0.04 \\
  NGC 7626 & 0.986 & 0.11 & 0.05 & 0.11 \\
  NGC 1427 & 0.990 & 0.09 & 0.03 & 0.04 \\
  NGC 1439 & 0.949 & 0.26 & 0.09 & 0.13 \\
  NGC 4589 & 0.979 & 0.13 & 0.05 & 0.06 \\
  NGC 5813 & 0.961 & 0.09 & 0.02 & 0.04 \\
  IC 1459  & 0.953 & 0.13 & 0.02 & 0.06 \\
\hline 
\end{tabular}
\end{table} 

\begin{figure*}
\vspace{1pt}
\hspace{10pt}
\epsfxsize=400pt
\epsfbox{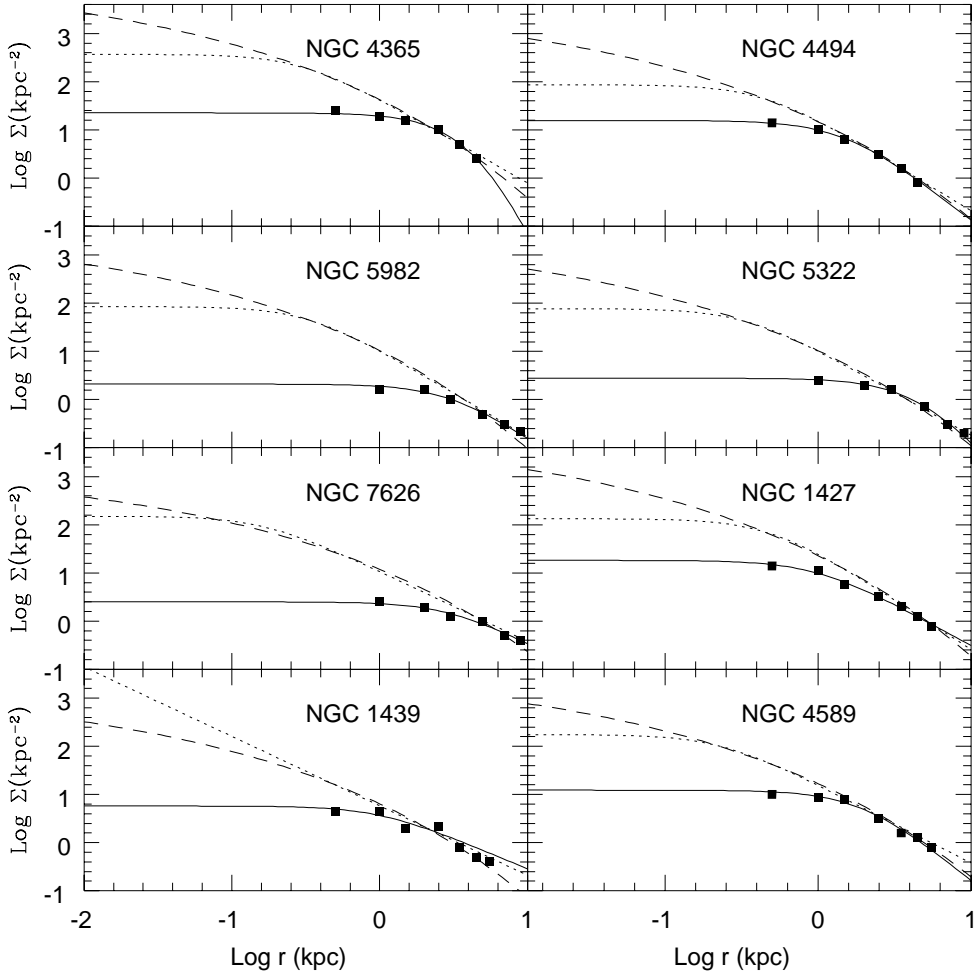}
\caption{Surface number densities for eight out of eleven galaxies of
our selected samples. Black squares represent the globular clusters observed
distribution, the solid line is its modified core model fit. Dashed and
dot-dashed lines are de Vaucouleurs' and modified core model fits to the
normalized (see text) galaxy profile, respectively.}
\end{figure*}
\section[]{The data and the results}
In this Section we discuss our analysis of the sample of globular cluster 
systems of Forbes et al. (1996).
These data constitute the first suitable, abundant, homogeneous set 
for globular clusters in galaxies taking advantage of the excellent resolution of the HST, which allows to determine the globular cluster content of galaxies in their inner regions,
where, indeed, it is expected to see the strongest signature of some formation and destruction processes. 

All the galaxies in the sample are ellipticals with kinematically distinct cores (KDC): this 
means that these galaxies have central regions that rotate rapidly, contrarily to the stars in the outer parts of the galaxy. It seems (Forbes et al., 1996) that kinematically distinct core ellipticals offer an ideal sample to study 
their GC population, in the mean time the results for this type of ellipticals  may also be applicable to "kinematically normal" galaxies, because it was shown (Forbes et al. 1995) that KDC
ellipticals follow the same scaling relations.

We selected galaxies in the sample according to the constraint
that the shape of the GCS surface density profile matches closely the 
profile of the galaxy halo light, at least in the outer regions. This is 
needed because we are making the case in which the globular clusters formed 
following the same spatial distribution of the halo stars.

According to this prescription, we are left with eleven out of
the fourteen GCSs, i.e.  of the following galaxies: 
NGC 4365, NGC 4494, NGC 1427, NGC 1439, NGC 4589, NGC 5813, 
IC1459, NGC 1700, NGC 5322, NGC 5982 and NGC 7626.
For every galaxy in this sub-sample, we made the following steps: 

$i)$ fit to the surface density profile of the GCS with a modified core model:
\begin{equation}
\Sigma(r) = {\Sigma_0 \over {(1+({r \over {r_c}})^2})^ \gamma}
\end{equation}
where $\gamma$ is a free parameter;

$ii)$ two analytical fits (de Vaucouleurs' and core model) to the 
light surface profiles of the galaxies, 
which have already been normalized by Forbes et al. to match the GCS distribution in the outer parts.

Since we are interested in the total number of clusters, which 
are connected to the 
surface integral of the curves, we obtained our fitting curves 
by minimising the usual least square loss function multiplied by 
the weighting factor $s$ ($s$ is the radial surface coordinate). 

\begin{table}
\caption{Total present number of clusters (second column), its initial value obtained with a de
Vaucouleurs model (third column), its fractional variation (fourth column) and the mass lost. Values are
obtained integrating on the range of available data points.
The colon indicates uncertainties in the results (see Sect. 2).}
\label{symbols}
\begin{tabular}{@{}lcccc}
Galaxy & $N$ & $N_i$ & $\Delta_{de V.}$ & $M_l$ ($M_{\odot}$)\\
\hline
  NGC 4365 & 502 & 683 & 0.27 & 3.71 $\cdot 10^7$ \\
  NGC 4494 & 190 & 240 & 0.21 & 1.01 $\cdot 10^7$ \\
  NGC 1700 &  23 & 26:  & 0.12: & 6.27 $\cdot 10^5:$ \\
  NGC 5322 & 166 & 187 & 0.11 & 2.06 $\cdot 10^6$ \\
  NGC 5982 & 128 & 177 & 0.28 & 5.0  $\cdot 10^6$ \\
  NGC 7626 & 208 & 287 & 0.28 & 1.61 $\cdot 10^7$ \\
  NGC 1427 & 235 & 392 & 0.40 & 3.15 $\cdot 10^7$ \\ 
  NGC 1439 & 127 & 127: & 0: & 0:\\
  NGC 4589 & 234 & 311 & 0.25 & 1.56 $\cdot 10^7$\\
  NGC 5813 & 375 & 511 & 0.27 & 2.77 $\cdot 10^7$\\
  IC 1459  & 264 & 415 & 0.36 & 3.00 $\cdot 10^7$\\
\hline
\end{tabular}
\end{table}
\begin{table}
\caption{As Table 2, but values are obtained extrapolating the fits to the galactic centre. The colon indicates uncertainties in the results (see Sect. 2).}
\label{symbols}
\begin{tabular}{@{}lcccc}
\\ Galaxy & $N$ & $N_i$ & $\Delta_{de V.}$ & $M_l$ ($M_{\odot}$)\\
\hline
  NGC 4365 & 517 & 849 & 0.39 & 6.65  $\cdot 10^7$\\
  NGC 4494 & 200 & 297 & 0.33 & 1.97  $\cdot 10^7$\\
  NGC 1700 &  25 & 39: & 0.36: &  2.82  $\cdot 10^6:$\\ 
  NGC 5322 & 175 & 266 & 0.34 &  1.82  $\cdot 10^7$\\
  NGC 5982 & 135 & 260 & 0.48 &  1.25  $\cdot 10^7$\\ 
  NGC 7626 & 215 & 365 & 0.41 &  3.01  $\cdot 10^7$\\
  NGC 1427 & 248 & 487 & 0.49 & 4.80  $\cdot 10^7$\\
  NGC 1439 & 130 & 141: & 0.08: & 2.20    $\cdot 10^6:$\\
  NGC 4589 & 241 & 371 & 0.35 &  2.61  $\cdot 10^7$\\
  NGC 5813 & 382 & 596 & 0.36 & 4.31 $\cdot 10^7$ \\ 
  IC 1459  & 271 & 516 & 0.47 & 4.87 $\cdot 10^7$\\
\hline 
\end{tabular}
\end{table} 
Assuming an analytical fitting curve for the galaxy light projected profile (vertically scaled to overlap the present GCS distribution in the outer, unevolved, regions) and another fitting curve to the observed GCS distribution, we can indeed compute the number of \lq lost\rq ~ clusters as difference
of these two surface distributions. In doing this, we follow the procedure outlined and applied to M87 first by McLaughlin (1995)  and subsequently discussed and improved by CDV who applied it to our Galaxy and M31, as well. 

The quality of the fits can be evaluated by Figs. 1 and 2 
as well as by the entries 
in Table 1, where we list the values of the correlation coefficients for the various fits and the loss function:
\begin{equation}
l = \sqrt{{{\Sigma(f_i-f_{0i})^2}s_i \Delta s_i} \over {\Sigma f_i^2 s_i \Delta s_i}}
\end{equation}
where $f_i$ and $f_{0i}$ are, respectively, the fitted and the observed value of the surface density in the annulus centred at the radial point $s_i$, of width $\Delta s_i$. It is easily seen that, in the majority of the cases, the analytical
fits reproduce well both the GCS data and
the galaxy light. 

With regard to the extrapolation of the analytical curves towards radii larger than the outermost original data
point, we note that the fitting model of the GCS and the bulge distribution may differ considerably in some cases,
but this has not any role on our results, for we perform the surface integrals
over a radial interval extending up to the outermost observed radial point.

In Table 2 we show the main results of our work:
the present and initial number of clusters ($N$ and $N_i$), $N_i$ being obtained (as described
before) integrating over the range of available data points.
The fraction of lost clusters to the total, $\Delta = (N_i-N)/N$,
is also given (here we list only the values obtained with the de Vaucouleurs' fit; the results for the core model fit differ only for a few percent).

\begin{figure*}
\vspace{1pt}
\hspace{25pt}
\epsfxsize=400pti
\epsfbox{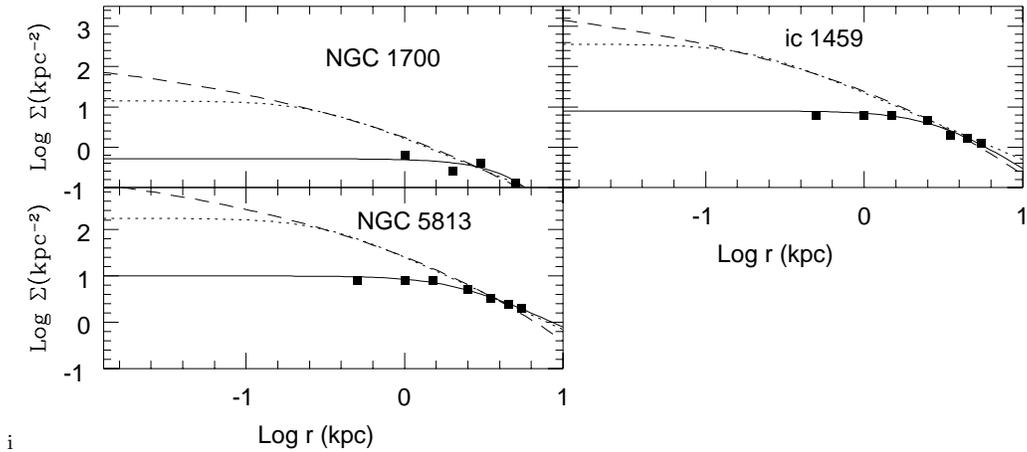}
\caption{Same as Fig. 1 for the remaining three galaxies of our sample.}
\end{figure*}

The fraction of lost clusters ranges from 0.10 to 0.40 of the total. 
Taking as average cluster mass, $\it<m>$, the Milky Way value of $2 \cdot 10^5 M_{\odot}$ (as obtained from data in Webbink 1985, assuming $(M/L)_{V,\odot}=1.5$) the mass lost by the GCS during its evolution can be easily computed.
We obtain, for the galaxies in our sample, an average mass lost  1.8 $\cdot 10^7 M_{\odot}$.
\par It is worth noted that in the case of NGC 1439 and NGC 1700, the fits are not good enough to give reliable results: in the first case this is probably due to the peculiar surface distribution of the GCS; in the second case to the very small number of clusters.
\par\noindent Extrapolating the fits to the centre we obtain the results of Table 3.
Here the fraction of lost clusters ranges from 0.33 to 0.49 of the present 
number, and the average mass lost is $3.2 \cdot 10^7 M_{\odot}$.
It is worth noting that the value of  $\it<m>$ $=2 \cdot 10^5 M_{\odot}$, being
the present average cluster mass in our Galaxy, is not necessarily 
a good representative of the actual, time-dependent value of the the mass of clusters falling 
to the galactic centre. By the way, as shown in CDT, the average cluster mass
in an evolved GCS is expected to vary less than one order of magnitude, for the
evolutionary effects (dynamical friction and tidal disruption) are at work on both sides
of the GCS IMF.

\section[]{GCS core radii and galactic mass correlation}
 In Table 4 we show the initial and the final values of the core
 radius of the GCSs, for each galaxy in the sample. This is obtained by fitting the initial and the final surface density  distributions with a modified Hubble law ($\gamma=1$ in Eq. 1):
\begin{equation}
\Sigma(r) = {\Sigma_0 \over {1+({r \over {r_c}})^2}}.
\end{equation}
We stress that while the initial core radii are very similar, their
final values differ significantly, as is clear from their ratios (third column of Table 4).

At this regard, we mention Fig. 10 of Forbes et al. 1996 which shows some correlation between
the GCS core radius and the parent galaxy luminosity: brighter (i.e. more massive)galaxies host more extended GCSs.
Forbes et al. interpret this observational result as a point against the role of
dynamical friction as an explanation for the observed radial distributions.

We disagree on this, for the following reasons. 
\par\noindent It is indeed correct
to say that dynamical friction is almost independent of the overall galaxy mass; anyway it is important to recall that 
dynamical friction alone acts just to increase the globular cluster density
around the centre of a galaxy, rather than producing any flat core.
Actually, CDT showed that it is the combined effect of dynamical friction and
tidal disruption of clusters  in the inner region of a galaxy that produces a rather flat GCS core. The time evolution 
of the GCS core radius is very dependent on the nucleus mass (see Fig. 6 of CDT)
because the tidal interaction with the central nucleus is, of course, stronger
for more massive nuclei, that deplete more effectively the GCS inner population.
It is so logical to expect a large GCS core radius in a brighter galaxy,
on the basis of the increasing evidence of a correlation between black hole mass and
parent galaxy mass (see, for instance, Kormendy \& Richstone 1995
and  Franceschini, Vercellone \& Fabian 1998).

\section[]{Conclusions}
The analysis of  the radial distributions of globular cluster systems of a sample
of elliptical galaxies drawn from data taken with the Hubble
Space Telecope WFPC2 by
Forbes et al. (1996) confirms  the already known result 
(e.g. Harris 1986, Capuzzo Dolcetta \& Vignola 1997) that globular cluster systems are always less concentrated to the centre than field halo stars.
Under the assumption that the initial GCS distribution and that of the halo stars
were initially similar in shape, we followed the procedure outlined in Capuzzo--Dolcetta
and Vignola (1997) to determine the initial number of clusters present in the galaxies  of the Forbes et al.  sample as the integral of the difference of the two normalized surface distributions.

We obtained that the percentage of clusters disappeared due to evolution
ranges, in the set of galaxies studied here, from about $30\%$ to $50\%$ of the present number. This implies
that a significant quantity of mass has been lost to a region around the centre 
of the parent galaxy, where the destruction mechanisms are particularly effective. The precise amount depends of course on the values of the initial individual globular cluster masses, which are difficult to be determined and  may vary from galaxy to galaxy.
Anyway,  assuming as cluster mean mass value $<m>=2\cdot 10^5$ M$_\odot$, as in the Milky Way,  
the amount of stellar mass lost to the environment is always above $1.25 \cdot 10^7 M_\odot$. 
This could explain the formation and activity of compact nuclei
in galaxies, as sketched in Capuzzo-Dolcetta (1993), Capuzzo-Dolcetta \& Tesseri (1997), Capuzzo-Dolcetta (1998). 

As a final result, the increasing core radius of the GCS surface distribution of brighter galaxies is explained in terms of evolution: the tidal interaction with the central compact
object is more efficient for more massive nuclei residing, usually, in more
massive galaxies.

\begin{table}
\caption{Observed core radii, their initial values and their ratios.}
 \label{symbols}
 \begin{tabular}{@{}lccc}
  Galaxy & $r_c$ & $r_{c0}$ & $ r_c / {r_{c0}}$ \\
\hline
  NGC 4365 & 1.8 & 0.4 & 4.5 \\
  NGC 4494 & 1.2 & 0.4 & 3   \\
  NGC 1700 & 2.4 & 0.5 &  4.8 \\
  NGC 5322 & 3.1 & 0.4 & 7.75 \\
  NGC 5982 & 3.3 & 0.4 & 8.25 \\
  NGC 7626 & 3.3 & 0.4 & 8.25 \\
  NGC 1427 & 1.2 & 0.5 & 2.4 \\
  NGC 1439 & 1.7 & 0.3 &  5.7\\
  NGC 4589 & 1.8 & 0.3 & 6\\
  NGC 5813 & 2.9 & 0.5 & 5.8 \\
  IC 1459  & 2.8 & 0.4 & 7 \\
\hline 
\end{tabular}
\end{table} 
\section*{Acknowledgments}

\appendix

\bsp

\label{lastpage}

\end{document}